\documentclass[pageno]{jpaper}
\pdfoutput=1
\usepackage[normalem]{ulem}
\usepackage{amsmath}

\usepackage{color}
\definecolor{orange}{rgb}{1,0.5,0}
\graphicspath{ {./graphics/} }

\begin{document}

\title{Sphynx: A Shared Instruction Cache Exporatory Study}

\author{\vspace{.08in} Dong-hyeon Park \and Akhil Bagaria \and Fabiha Hannan \and
  Eric Storm \and Josef Spjut}

\date{}
\maketitle

\thispagestyle{empty}

\begin{abstract}
The Sphynx project was an exploratory study to discover what might be
done to improve the heavy replication of instructions in independent
instruction caches for a massively parallel machine where a single
program is executing across all of the cores. 
While a machine with only many cores (fewer than 50) might not have
any issues replicating the instructions for each core, as we approach
the era where thousands of cores can be placed on one chip, the
overhead of instruction replication may become unacceptably large.
We believe that a large amount of sharing should be possible when the
machine is configured for all of the threads to issue from the same
set of instructions.
We propose a technique that allows sharing an instruction cache among
a number of independent processor cores to allow for inter-thread
sharing and reuse of instruction memory.
While we do not have test cases to demonstrate the potential magnitude
of performance gains that could be achieved, the potential for sharing
reduces the die area required for instruction storage on chip.
%% We propose sharing a single instruction cache among a group of threads
%% by providing independent reads as long as the threads access different
%% cache banks.
%% \jbs{Need to add a better summary of the proposed technique once it's
%% written.}

\end{abstract}

\section{Introduction}

%% \jbs{I expect this entire paper to be about 3-4 pages. We should be clear
%% and concise, but still put all our ideas down.}

Instruction caches are widely used to mediate the effects of reads
from main memory, relative to computation time. 
The instruction cache is accessed for every instruction executed and
program execution time can vary widely depending on the number of
instruction cache misses~\cite{arnold94}. 
In existing Graphics Processing Units (GPUs) and CPUs, each processor
core has its own instruction cache. 
A unified or shared instruction cache that is used by many or all
cores of a GPU or CPU has the potential to improve system performance
and reduce power consumption.
However, such a modification also results in increased traffic for the
instruction cache, which could lead to a higher miss rate, reducing
performance and increasing power consumption. 

%seems like a contradiction between the above (higher miss rate) and below (reduced compulsory miss rate)
%We may want to clarify...

In current computer architectures, the instruction caches are
independent of one another. 
However, since each processor uses the same
set of instructions, it is plausible that a shared instruction cache
could introduce non-trivial improvements in performance. 
Unified instruction caches could reduce the number of compulsory
misses because an instruction previously executed by one streaming
multiprocessor may be available for another streaming multiprocessor
immediately rather than after an additional miss is serviced. 
In addition to a fully unified instruction cache used across all
processor cores, another possible solution could be to maintain
multiple instruction caches, which are shared across a subset of all
cores. 
This architecture could allow the operating system to group together
program threads based on similarity of instructions to maximize the
benefit of shared instructions and minimize the conflicts across
threads.

\section{Background}

An instruction cache can have a large impact in a processor's
performance. 
There has been lot of work in the past in improving the performance
of instruction cache  CPUs. 
Techniques such as advanced branch prediction~\cite{yeh93} and
replacement policies~\cite{smith85} have contributed to the high
performance of instruction caches in modern CPUs. 
It has even been shown that system performance can increase by 10-20\%
just by adjusting the operating system to use the instruction cache
efficiently~\cite{torrellas98}.
While all of these approaches are useful, the future of computing
includes very large numbers of independent processor cores integrated
on one chip.
While general purpose CPUs are likely to include hundreds of cores at
some point in the future, a good example to consider today is the GPU,
which already exposes hundreds to thousands of threads to the
programmer. 

Modern GPUs are designed to have several
compute units to maximize throughput, and to expose these compute
units for general purpose computation instead of only graphics
applications.
These compute units are compact, and designed to perform simple
operations on a large set of data. 
It is not feasible to implement techniques such as advanced branch
prediction on these small compute units, because a single GPU unit
can have hundred of these compute units. 
To improve the instruction cache performance of GPUs, new solutions
need to be developed in ways that maintain the compact and simplistic
nature of the compute units. 
Current GPU architectures are designed with individual L1 instruction
cache for each compute unit of the GPU\cite{keckler11}. 
It may be possible to have multiple compute units share the same
instruction cache without incurring serious performance or power
penalties.

\section{Shared Instruction Cache Design}
%\jbs{Akhil in charge.}

%% \begin{figure}[ht!]
%% \centering
%% \includegraphics[width=\columnwidth]{graphics/InstructionCacheDesignSketches.jpg}
%% \caption{(a) Current design of Instruction Caches in an SM, (b) Proposed cache design 1 (b) Proposed cache desgn 2.}
%% \label{propDesign}
%% \end{figure}

\begin{figure}
\centering
\includegraphics[width=\columnwidth]{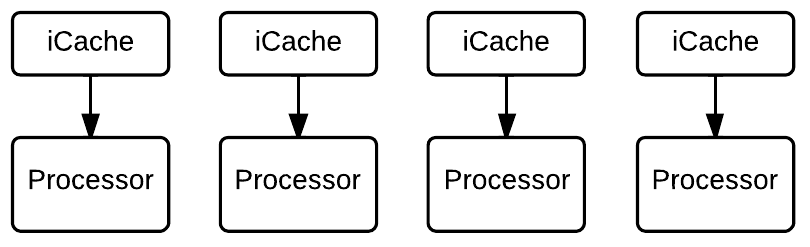}
\caption{Independent instruction caches}
\label{fig:indep}
\end{figure}

Most computer architectures build an independent instruction cache for
each processing core as seen in Figure~\ref{fig:indep}.
Reducing the number of instruction caches for the same number of cores
in a multi-core processor may have certain advantages. 
First, a shared instruction cache design could reduce the amount of
storage and die area required to hold instructions on chip when
multiple processors are executing the same program.
This advantage in storage and area could result in reduced performance
if the pressure to this shared resource becomes too high.
Similarly, the benefits of reduced cache size will not be realized in
multi-program workloads, unless the operating system is able to
discover shared library code and perform the proper virtual mapping to
allow the hardware to exploit it.
Second, a shared instruction cache design could reduce the number of
compulsory misses because an instruction previously fetched
by one processor may be available for another processor immediately
rather than requiring an additional cache miss. 
Again, this benefit is only relevant for single-application parallel
workloads. 

\begin{figure}
\centering
\includegraphics[width=\columnwidth]{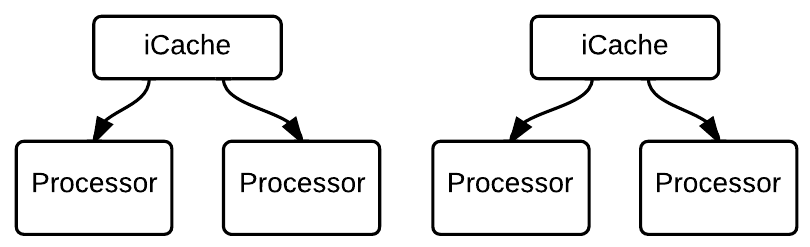}
\caption{Partly shared instruction caches}
\label{fig:paired}
\end{figure}

\begin{figure}[ht!]
\centering
\includegraphics[width=\columnwidth]{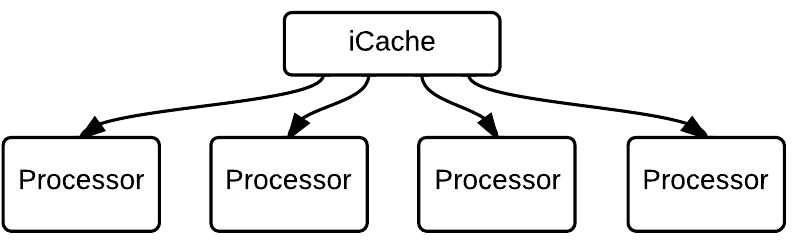}
\caption{Fully shared instruction caches}
\label{fig:shared}
\end{figure}

The first approach to shared instruction caches would be to couple an
instruction cache with a pair of processors as in
Figure~\ref{fig:paired}. 
This would require a small amount of routing overhead and allow for the
operating system or programmer to intelligently schedule two threads 
that share instructions to these two processor cores in order to exploit 
the shared instruction cache.
While sharing an instruction between two cores is one step towards shared
caches, at its limit, a chip could be designed to share a single instruction
cache among all processors on chip, shown with an example of only four
processors in Figure~\ref{fig:shared}.
For an arbitrary number of cores, the level of sharing could be
scaled up or down to fit the applications and purpose of the
architecture design.
One could even imagine a design where assymetrical sharing is enabled
by grouping different sets of compute cores differently and then
assigning the application threads to the appropriate cores to match
the sharing to the hardware available.

While our proposed technique is likely only beneficial in situations
with highly parallel programs, it is probable that future systems will
often run a few highly parallel workloads that can benefit from these
advantages. 
A hybrid approach would allow some cores to efficiently process general 
purpose applications while cores with shared instruction caches execute
parallel single program workloads.
When different threads within a parallel application diverge in program flow,
it may be useful to further design the instruction caches to have multiple
independent banks that can be accessed in parallel.
We believe the multi-banked approach to be the most beneficial, however we do not 
present results or analysis of multi-banked caches in this work.

\section{Results}

% expected results
There are two methods for analyzing the potential gains from shared
instruction caches.
The first is to keep the total instruction cache size across the chip
constant and increase the capacity each thread sees by grouping that
storage together.
This method of sharing will provide no area savings, but should grant
an increased hit rate for the now larger cache.
The increase will be limited by the working set of the application.
We expect that the hit rate improvement will only be minor and
therefore do not generate any results for this method.
%% We project improvement in cache performance for this approach in
%% Figure~\ref{HitImprov}. 

% \begin{figure}[b]
% \centering
% \includegraphics[width=\columnwidth]{graphics/HitRateImprov.png}
% \caption{Projected instruction cache hit rate with fixed total cache size and increased sharing}
% \label{HitImprov}
% \end{figure}

\begin{figure}
\centering
\includegraphics[width=\columnwidth]{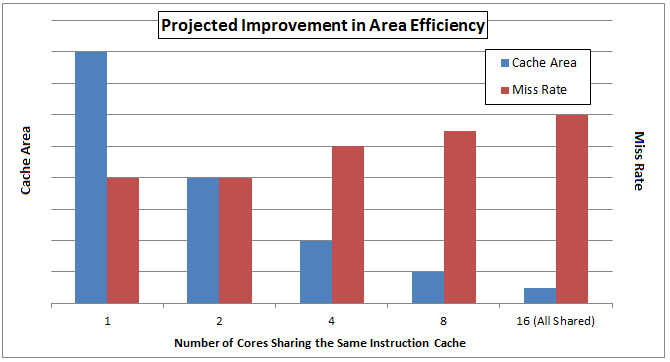}
\caption{Hypothesis of area and miss rate for shared caches}
\label{AreaEff}
\end{figure}

The second method is to reduce the total storage required for
instruction cache data arrays while allowing at most a minor
degredation in hit rate.
We consider an approach where the instruction cache each processor
sees remains fixed but the number of threads sharing each cache is
increased. 
We expect that the miss rate of the cache will increase when the instruction
cache is not large enough to satisfy the demands from all the cores
sharing the cache.
However, the performance penalty should not be as dramatic as the
reduction in area when all threads run the same application.
Figure \ref{AreaEff} shows our hypothesis for this approach
which should improve the efficiency of the chip.
Note that the scale on the y-axis is omitted because the chart is included
to build intuition of the hypothesis only and is not based on 
any simulation or otherwise realistic result.

\subsection{Setup}

We use GPGPU-Sim~\cite{bakhodayuan09} to analyze the effectiveness of our design because of
its ability to simulate a large number of parallel processing cores.
GPGPU-Sim is an open-source software package
available to simulate GPU architecture. 
It has been validated to be representative of performance on NVIDIA
GPUs and provides a reasonable platform for testing alternate
highly-parallel computer architectures. 
We use a reference configuration for a NVIDIA GTX580 GPU for our study.
The GTX580 contains 16 streaming multiprocessors (SM) with the NVIDIA Fermi
architecture. 
In GPGPU-Sim, each CUDA streaming multiprocessor is represented as a
single SIMT core, with all the SIMT cores placed within a single SIMT cluster. 
Each streaming multiprocressor can have up to 48 warps, with 32
threads per warp (See Table \ref{table:gpuconfig}). 
All sixteen SIMT cores share unified 786KB L2 cache. 

% \begin{figure}[b]
% \centering
% \includegraphics[width=\columnwidth]{graphics/GTX580.jpg}
% \caption{Block Diagram of NVIDIA GTX580~\cite{gf100}}
% \label{GTX580}
% \end{figure}

\begin{table}[ht]
\caption{GTX580 Configuration in GPGPU-Sim}
\begin{tabular}{l|c}
\hline\hline
SIMT cluster count & 1 \textbf{/} 2 \textbf{/} 4 \textbf{/} 8 \textbf{/} 16  \\
\hline
Cores per cluster & 1 \\
\hline
Memory controller count & 6 \\
\hline
Subpartition per mem. & 2 \\ 
\hline
Shader registers & 32768x(16 \textbf{/} 8 \textbf{/} 4 \textbf{/} 2 \textbf{/} 1 ) \\
\hline
Threads in pipeline & 1536x(16 \textbf{/} 8 \textbf{/} 4 \textbf{/} 2 \textbf{/} 1 ) \\
\hline
Threads per warp & 32 \\
\hline
Scheduler per core & 2x(16 \textbf{/} 8 \textbf{/} 4 \textbf{/} 2 \textbf{/} 1 )\\
\hline
CTA per core & 8x(16 \textbf{/} 8 \textbf{/} 4 \textbf{/} 2 \textbf{/} 1 )\\
\hline
Core clock & 700 MHz\\
\hline
L2 \& Interconnect clock & 700 MHz\\
\hline
DRAM clock & 924 MHz\\
\hline
Topology & 13 \textbf{/} 14 \textbf{/} 16 \textbf{/} 20 \textbf{/} 28 \\
\hline
L1 Instruction Cache & 4 sets : 128B blocks : 4-way \\
\bottomrule[1pt]
\end{tabular}
\caption*{GPGPU-Sim configurations used in simulation. 
Any parameter not listed in the table matches the original GTX580 configuration included in the public release of GPGPU-Sim.}
\label{table:gpuconfig}
\end{table}

In the Fermi architecture, each streaming multiprocessor has
its own distinct L1 cache. 
Each L1 instruction cache is 4-way set associative, with 4 sets of
128 bytes per block. 
To assess the  performance of different L1 instruction cache
architectures, only the L1 instruction cache was modified, while all
other architectural variables remained constant. 
Other aspects of the architecture such as the bandwidth of shared L2
cache could serve as a performance bottleneck for different cache
designs. 
However, these variables are ignored for this study, because we expect
them to have little impact on the hit and stall rates of the
instruction cache, which are the primary metrics of interest.

\subsection{Preliminary Results}

\begin{figure}
\centering
\includegraphics[width=\columnwidth]{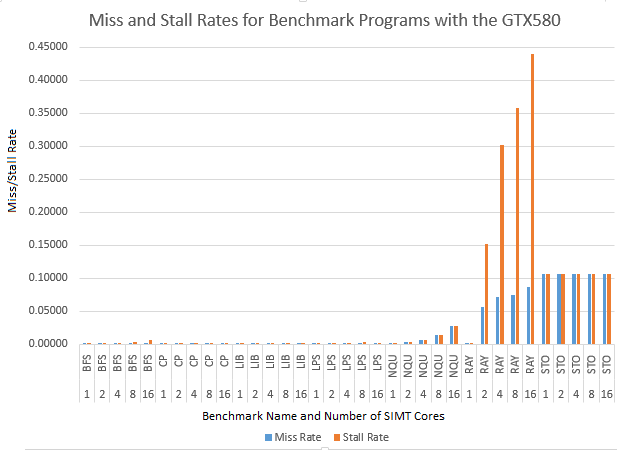}
\caption{Simulated miss and stall rates for configurations in Table~\ref{table:gpuconfig}}
\label{fig:missStalls}
\end{figure}

We used benchmarks provided by the standard GPGPU-Sim distribution
to validate our second method for sharing instruction caches by
keeping the cache the same but increasing the number of threads that
share that cache.
This means varying the number of SMs per instruction cache from 1 to
16 (the maximum number of cores available on the GTX580).
Note that since each SM can have up to 48 warps, the number of
simultaneously executing threads sharing the same instruction cache is
much larger than the SM count.
The results can be seen in Figures~\ref{fig:missStalls}
and~\ref{fig:missStallsZoomed}. 
As expected, the miss rate increases slightly in most cases by the
increased pressure on the shared cache.
A couple of benchmarks were more problematic, resulting in
almost a 45\% stall rate when shared by 16 threads.
We provide observations about those benchmarks in
Section~\ref{sec:benchmarks}.

For the cases where the miss rate becomes unacceptable under increased
sharing, we should allow for the cache to increase in
capacity. 
However, capacity alone will only affect the miss rate reported from
the simulations.
The stall rate represents the percentage of cache accesses that fail
due to the increased pressure on the cache from sharing without
increasing the available read parallelism.
We expect that simply allowing for independent banks of the
instruction cache to be accessed independently would overcome this
limitation, but we have not yet simulated this approach.
However, other high-level simulation has shown this kind of
instruction cache banking scheme to be potentially
effective~\cite{kopta10}.

\subsection{Benchmarks}
\label{sec:benchmarks}
\begin{figure}
\centering
\includegraphics[width=\columnwidth]{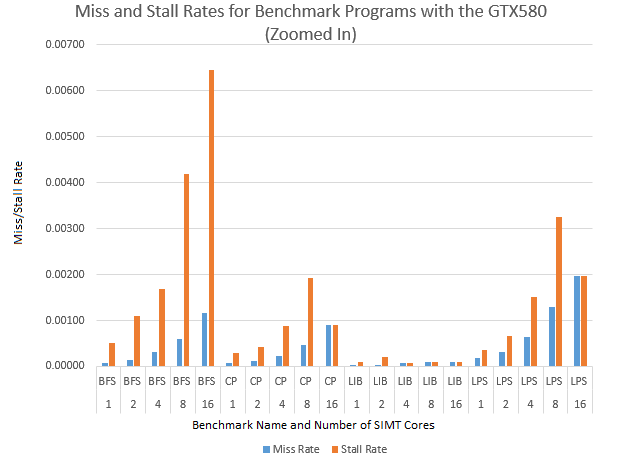}
\caption{Simulated miss and stall rates (zoomed-in Figure~\ref{fig:missStalls}) }
\label{fig:missStallsZoomed}
\end{figure}

The result of varying the number of SM sharing from 1 to 16 is shown in 
Figure~\ref{fig:missStalls}, and Figure~\ref{fig:missStallsZoomed} shows 
a zoomed-in version of the same data to show that even when the magnitude 
of misses is very small, the expected trend is followed. 
A total of seven benchmarks were
tested, each available from the GPGPU-Sim source code \cite{bakhodayuan09}.

The benchmark STO is mostly unaffected by the increase in number of cores
accessing the same small instruction cache, and the benchmark shows a similar
level of miss and stall rate. 
On the other hand, RAY exhibits a much higher stall rate than its miss rates in Figure~\ref{fig:missStalls}, 
and is evidently more sensitive to increased sharing of the cache.
Higher stall rate compared to miss rate occurs when the cache is not able to 
respond to the CPU on time, even when there is a cache hit.
The rapid increase in stall rate in RAY and the relatively modest increase in miss rate 
suggests the cache is not able to handle the rapid increase in the amount of requests coming from the cores,
and has to stall even when there is a hit.

As RAY showed to have the largest number of PTX instructions amongst our set of benchmarks,
the high miss and stall rate is reflective of the high level of instruction cache utilization
by the benchmark. 

Similar relationships between stall rate and miss rate are also evident in 
benchmarks BFS, CP, and LPS. However, the miss rate of less than 1\% in these benchmarks
suggests that the benchmarks are largely affected by initial compulsary misses, and operate
largely on a small set of instructions.

% \emph{CP (Coulombic Potential)}

% Given a standard configuration, CP issues a high number of thread
% blocks -- a representation of the high degree of parallelism that CP
% utilizes.

% \emph{Ray (Ray Tracer)}

% RAY was the most interesting bechmark because it had the highest
% number of PTX instructions issued and was also significatly parallel. 
% Moreover, the fact that the degree of parallelism of RAY could be
% increased easily by modifying the parameters to the CUDA program.
% By increasing the size of the picture to 512 by 512, we made RAY issue
% a much higher number of thread blocks per unit time. 
% This ensured that the RAY benchmark was executing as parallel in
% software as the Coulombic Potential benchmark in the number of work
% elements that it was issuing. 
% Given that the RAY CUDA file issues more instructions in its core
% computation that CP does, and that a lot of its instructions (CUDA
% instructions) seem to be computationally more expensive, RAY can now
% be treated as both a long program which would occupy a significant
% amount of space in the instruction cache (and hence would not be
% entirely resident in the instruction cache) AND would issue those
% instructions with a high degree of parallelism.

\section{Conclusions}

The proposed instruction cache designs provide a possible solution to 
increase the efficiency of instruction cache in parallel processors, 
in particular processors whose primary workload includes single programs
with many parallel threads executing the same code. 
The sharing of instruction cache amongst multiple cores can help 
chip designers optimize for less area without compromising performance. 
The extra space recovered from the sharing of instruction cache among 
multiple cores can be repurposed to improve other parts of the 
chip, such as a larger data cache(s) or additional computation units.
We showed results for GPU workloads because they typically exhibit much 
higher levels of parallelism than CPUs while still executing a single 
application.
GPUs traditionally only expose multi-program workload capabilities 
using corse-grained time-sharing among processes.

For future work, the proposed instruction cache design should be 
simulated and tested to verify the affectiveness of the design. 
The different cache architecture parameters that are expected to 
affect the performance of instruction cache design are: number of 
cores sharing the cache, associativity and size of the cache. 
Extensive testing of such parameters should be conducted to determine 
the most optimal instruction cache design for common workloads. 
The proposed design is expected to perform the best on multi-threaded 
applications with a large amount of redundant instructions, so the 
drawbacks of running non-redundant code should also be considered.

\section*{Acknowledgement}
We acknowledge that this idea for reducing the instruction cache area
overhead by aggressively sharing the cache originally came from
previous work on the TRaX
architecture~\cite{spjut08,spjut09,kopta10,spjut12,kopta13,kopta14}.
The TRaX architecture was designed for ray tracing and supports
thousands of threads running the same application but working on
different data.
We would also like to acknowledge the reviews from Konstantin Shkurko and Steven Jacobs.
This work was completed at Harvey Mudd College in Claremont, California.

\bstctlcite{bstctl:etal, bstctl:nodash, bstctl:simpurl}
\bibliographystyle{IEEEtranS}
\bibliography{references}

\end{document}